# Chemical bonding origin of the thermoelectric power factor in Half-Heusler semiconductors

Kasper Tolborg[a] and Bo B. Iversen[a]*

a) Center for Materials Crystallography, Department of Chemistry and iNANO, Aarhus University, Langelandsgade 140, 8000 Aarhus C, Denmark
Corresponding author: bo@chem.au.dk

**Abstract**

Intermetallic semiconductors with the cubic Half-Heusler structure (XYZ) have excellent thermoelectric properties. This has been attributed to the high degeneracy of the carrier pockets in the band structure, but large differences are found between different material compositions. Half-Heuslers are often interpreted within Zintl chemistry, making a clear distinction between an electropositive cation ($X^{n+}$) and an extended polyanion ($YZ^{n-}$). Based on quantitative real space chemical bonding analysis, we unravel large degrees of covalent bonding between the formal cation and anion, making the Zintl distinction clearly invalid. This covalence is shown to strongly affect the band structure, thermoelectric properties and response properties in the materials, with improved thermoelectric properties observed for those materials that least follow the Zintl concept. This expands our knowledge of the chemical bonding motifs governing physical properties, and gives a critical view on the simplistic chemical concepts too often applied for design of complex materials.

## 1. Introduction

The world is in high demand for development of sustainable energy sources in order to mitigate the impact of the global climate crisis. A large amount of energy is wasted as heat, and thus a promising method for decreasing our energy consumption and greenhouse gas emissions is through waste heat harvesting. Thermoelectric materials are able to interconvert thermal and electrical energy, and thus offer a possibility to achieve this goal through solid-state devices. Currently, thermoelectric performance is limited by the efficiency of the materials, which is determined by the dimensionless figure of merit $zT = S^2\sigma T/\kappa$, where $S$ is the Seebeck coefficient, $\sigma$ is the electrical conductivity, and $\kappa = \kappa_L + \kappa_e$ is the thermal conductivity divided into lattice and electronic parts.[1] These parameters are highly counteracting, which has led to the introduction of the phonon-glass electron-crystal concept, since the electronic parameters are optimized for crystalline solids, whereas disorder decreases the thermal conductivity.[2] However, even the electronic parameters governing the power factor, $S^2\sigma$, counteract each other, since the Seebeck coefficient is optimized for low carrier concentrations and heavy electronic bands, whereas the electrical conductivity is increased for large carrier concentrations and light bands. Design strategies for optimizing the power factor thus often rely on finding and engineering materials with degenerate or anisotropic carrier pockets.[3-8] Since the design of thermoelectric materials is highly challenging, there is a need for developing an understanding of the relations between chemical bonding and thermoelectric properties, although this relation is often far from straightforward.[9-11] The inaccessibility of fundamental thermoelectric materials design has often led to application of very simplistic chemical concepts, but in order to avoid misguiding, it is important to understand the general validity of simple concepts for explaining the nature of the chemical bonds and their relation to material properties.

Half-Heusler (HH) compounds with the XYZ stoichiometry represent a promising materials class for high temperature thermoelectric applications.[12-14] The HH compounds crystallize in the cubic space group $F\bar{4}3m$, and they consist of a tetrahedral YZ zinc blende network with X in the octahedral voids, meaning that XZ forms a rock salt network, while XY also forms a zinc blende network (Figure 1a). Here all HHs will be denoted as XYZ with the Y atom being part of both zinc blende networks, and X being the most electropositive element. Thermodynamically stable stoichiometric HHs are valence-balanced semiconductors with valence electron counts of 8, if they consist solely of main group elements, or 18, if transition metals are present.[12] HHs have promising electronic properties for thermoelectric applications, which are suggested to originate in large band degeneracy and weak electron-phonon coupling, giving simultaneously large Seebeck coefficient and carrier mobilities.[15-18] However, the thermal conductivity of the pristine materials is too high for practical applications, and thus, considerable effort is spend towards reducing thermal conductivity through alloying and nano-structuring.[12, 14]

## 1.1 Zintl chemistry of Half-Heuslers

The valence balanced semiconducting nature has led to the interpretation of HHs as Zintl phases, and it has been suggested that Zintl chemistry can be used to engineer the properties of HHs.[13] In Zintl phases, a formal cation donates its valence electrons to a polyanion, which forms a valence balanced covalent bonding pattern fulfilling the octet rule.[19, 20] Thus, ionic and covalent bonding patterns coexist in these materials. Applying the Zintl concept to HHs, the $X^{n+}$ cation formally donates its valence electrons to a negatively charged covalently bonded $[YZ]^{n-}$ zinc blende network for which either the 8- or 18-electron rule is fulfilled.[13] Recently, also defect HH structures with a formal valence count of 19 have been discovered, but introduction of vacancies on the X site reduces the electron count to 18 in line with the Zintl concept.[21-23] Zintl chemistry is a desirable concept for engineering of thermoelectric materials, as it in principle allows for decoupling the electronic and thermal transport properties through increased carrier mobility in the covalent substructure and reduced thermal conductivity from the ionic substructure by introducing disorder.[1, 24, 25]

Besides predicting the stability of certain stoichiometries, the Zintl concept also gives rise to predictions regarding the electronic structure and defect chemistry of HHs as reviewed by Zeier *et al*.[13] In a pure Zintl picture, the band gap should be formed between the valence states of the polyanion and the valence states on the cation, but going beyond pure Zintl chemistry a further hybridization between d-states on X and Y is expected.[13] Regarding defect chemistry, Zintl chemistry gives an explanation for the observed large off-stoichiometries with interstitial Ni on the vacant tetrahedral site in XNiSn HHs,[26, 27] since a further subdivision of oxidation states suggests easy formation of neutral Ni interstitials.[13]

In this article, we critically examine the predictions from Zintl chemistry beyond the stability of certain electron counts based on density functional theory (DFT) calculations and chemical bonding analysis, and we use this to bridge the gap between the understanding of the electronic structure, thermoelectric properties and chemical bonding in HH semiconductors.

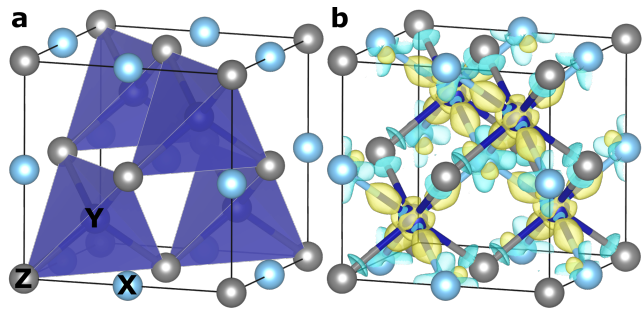

**Figure 1.** Crystal structure and deformation density in Half Heuslers. **a** Half Heusler crystal structure with the Y atom located at the unique position forming part of two zinc blende networks. **b** Deformation electron density, $\Delta\rho = \rho_{\text{DFT}} - \rho_{\text{IAM}}$ (IAM: independent atom model), of TiCoSb shown as a representative of the HHs. Ti is drawn in light blue, Co in dark blue and Sb in grey. Positive isosurfaces are shown in yellow and negative isosurface in blue. The isosurfaces are drawn at 0.01 e au$^{-3}$.

## 2. Results and discussion

### 2.1 Real space chemical bonding in Half-Heuslers

Zintl chemistry suggests that HHs consist of a covalently bonded [YZ]$^{n-}$ polyanion, and an X$^{n+}$ cation donating its valence electrons to the extended polyanion. This gives rise to a division into Li$^+$[SiAl]$^-$ and Ti$^{4+}$[CoSb]$^{4-}$ for typical main group and transition metal based HHs. Although one should not equate the oxidation state and ionicity in a material,[28] Zintl chemistry thus predicts highly dissimilar chemical bonding within the formal extended polyanion and between the formal cation and the polyanion.

Real space chemical bonding analysis based on descriptors derived from quantum chemistry allows us to analyze the validity of this chemical bonding scheme. Here we apply Bader's Quantum Theory of Atoms in Molecules (QTAIM)[29] and delocalization indices (DI)[30, 31] to assess the chemical bonding in HH materials. From QTAIM, space is partitioned into atomic basins, and the electron density is integrated within an atomic basin to yield the atomic charge, and the exchange-correlation density is integrated over two atomic basins to obtain the delocalization index, which is a quantum mechanical measure of the bond order, i.e. two times the DI gives the number of electrons shared in the chemical bond. Thus, evaluation of charge transfer and DIs allows us to assess the ionicity and covalency in a material. Bende et al.[32, 33] have previously applied these tools on HHs to explain why main group HHs crystallize with the electronegative late main group element on the Y position, whereas transition metal HHs crystallize with the late transition metal on the Y position.[32] Furthermore, they showed that bonding within the polyanion for the main group HHs is consistent with the octet rule and strongly polar covalent bonding as suggested by Zintl chemistry.[33] Wuttig et al. have furthermore used the concepts to build a map of chemical bonding in binary solids, which allowed them to identify a region of metavalent bonding with improved response properties for thermoelectrics and phase change materials.[8, 34]

The results of our chemical bonding analysis are shown in Table S1 in the supporting information and summarized in Figure 2 using a map of chemical bonding inspired by those of Wuttig et al.[8, 34] The chemical bonding situation expected from a Zintl perspective is approached with increasing charge transfer from the formal cation to the extended polyanion, and with a bond order approaching one within the Zintl anion and approaching zero between the cation and polyanion. This situation is approached in the bottom right corner of both panels in Figure 2. We observe that the Zintl perspective is reasonably fulfilled for main group based HHs, especially LiSiAl and LiAsZn, for which the charge transfer is close to one, and the bond order within the polyanion is close to or above 0.5, whereas it is less than 0.1 between cation and either elements from the polyanion. Although the Zintl concept is reasonably fulfilled here, the bond order is significantly less than one, which is the result of charge transfer within the polyanion. For LiAsMg, the charge transfer within the polyanion is larger, and the bond order lower than for the other main group systems as expected from the larger electronegativity difference. This is very similar to what we have previously shown for thermoelectric antimonides in the $CaAl_2Si_2$-type structure, where materials with Zn on the Al-site align with the Zintl concept, whereas those with Mg on the Al-site do not. In those cases, the chemical bonding differences led to differences in anisotropy of the physical properties.[11]

For the transition metal containing HHs, the bonding scheme is more complex. In all cases, the Zintl cation is positively charged, but significantly less than suggested by the formal oxidation state. Within the negatively charged polyanion, the charge transfer is, however, very counterintuitive, since the main negative charge is in almost all cases attained by the late transition metal rather than the more electronegative main group element. This adds to the group of reverse or small charge transfer observed in transition metal antimonides,[35, 36] which is at least partly captured by more recent electronegativity scales.[37] This suggests that late transition metals may be considered more electronegative than the late main group elements, opposite to what is commonly used. Regarding bond orders, we observe that they are generally highly similar within the Zintl anion and between the cation and anion. Thus, in terms of the degree of covalency, bonding is very similar within both zinc blende networks, and the large difference suggested by Zintl chemistry is not observed here. Similar differences are observed in terms of other bond indicators such as the electron density at the bond critical point reported in Table S1 and summarized in Figure S6 in the supporting information, and deformation densities as shown in Figure 1b for TiCoSb – both of which are experimentally accessible.[10]

Clearly, the concepts from Zintl chemistry cannot be aligned with a detailed real space analysis of chemical bonding in HHs. This is also reflected in the population of X d-states, which are formally empty in the Zintl picture, but hybridizes with the Y d-states to form the bands at the top of the valence band. The importance of the hybridization increases significantly with the covalency of the XY-interaction as seen in Table S1 and Figure S8 in the SI.

In addition to the overall bonding scheme, we observe significant differences between the different transition metal containing HHs, and as evident from Figure 2, the Zintl concept is approached for the group III-X-XV HHs (e.g. ScNiSb), whereas the chemical bonding deviates more from Zintl chemistry in the V-IV-XIV and V-VIII-XV systems (e.g. NbCoSn and VFeSb). Thus, the difference in main group number, and correspondingly electronegativity between the two transition metals is very important for the degree of covalent bonding, which is not surprising, but nevertheless in starch contrast to the often applied Zintl chemistry picture.

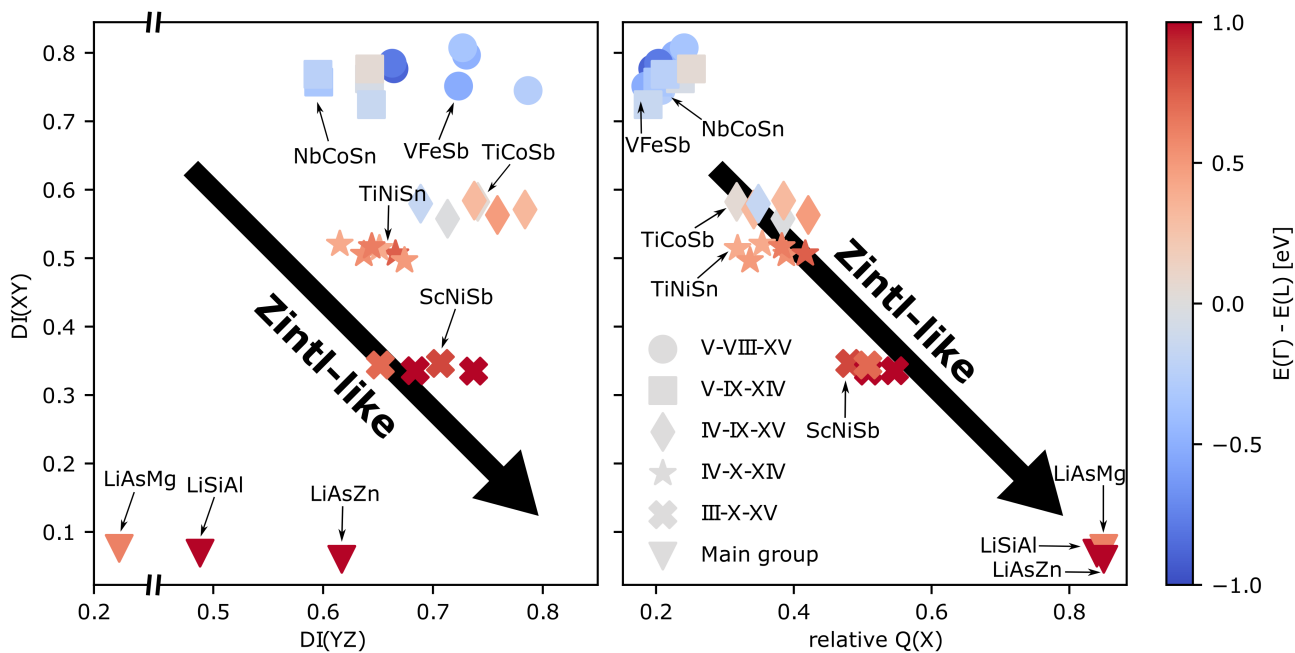

**Figure 2.** Maps of chemical bonding in Half Heuslers. Main group and transition metal containing HHs are marked according to their chemical bonding descriptors. On the ordinate the delocalization index (DI) between X and Y atoms (i.e. the non-Zintl zinc blende framework) is shown, and on the abscissa the DI between Y and Z (i.e. the Zintl zinc blende framework) is shown to left, and the charge transfer from formal Zintl cation to Zintl polyanion divided by the formal oxidation state is shown to the right. Arrows points towards the area with the expected behavior, if HHs are interpreted within the Zintl concept, i.e. weak covalent interaction between cation and anion, significant charge transfer from cation to anion, and significant covalency within the Zintl anion. The materials containing elements from the same groups in the periodic table are marked with the same symbols for the transition metal containing ones, and with one symbol for all main group HHs. The materials are colored according to the energy difference between the highest occupied states at the Γ-point and L-point in the first Brillouin zone. Numerical values are found in Table S1 and S2 in the SI.

## 2.2 Chemical bonding effect on thermoelectric properties

The discussion of whether or not Zintl chemistry is applicable to HHs is not purely of semantic nature. Rather, it has been noted in the literature that their carrier pockets in the valence band may occur at different high symmetry points in the first Brillouin zone depending on composition,[16, 18] leading to different valley degeneracies and therefore significantly different thermoelectric power factors for p-type materials. The valence band maximum (VBM) may occur at the Γ-point (with a valley degeneracy of 1), L-point (4) or W-point (6), or two or more points may be effectively converged.[16] In Figure 2, the materials are colored according to the energy difference between the highest occupied bands at the Γ- and L-points. It is evident that approaching Zintl chemistry in the lower right corner, the VBM is dominated by the Γ-point, whereas the L- and W-points dominates with increasing covalent interactions between the two transition metals.

The analysis can be expanded to take into account all contributions from the band structure to the thermoelectric power factor, which is maximized for high density of states effective mass and low conductivity effective mass. This can be achieved by increasing the carrier pocket degeneracies and anisotropies.[6-8] To obtain a measure of the band structure contribution to the thermoelectric power factor, Gibbs *et al.*[38] introduced the Fermi surface complexity factor, $N_V^* K^*$, which effectively gives the product of the degeneracy and the anisotropy of the Fermi surface. Figure 3a shows the Fermi surface complexity factor for a range of transition metal containing HHs mapped onto their chemical bonding descriptors. Again, we observe clear trends for materials with similar group elements to show similar properties, and we observe that the Fermi surface complexity factor is maximized for the systems with strong covalent interactions between the formal Zintl cation and polyanion. The optimal situation for increasing the thermoelectric power factor is obtained for strong covalency between the XY framework, and weaker covalency within the Zintl anion. This is the case for the V-IX-XIV HHs, e.g. NbCoSn, for which the L- and W-points are effectively converged at the VBM. However, the relationship is not purely that large XY covalency results in a larger Fermi surface complexity factor, as evident from a comparison of the IV-IX-XV systems, e.g. TiCoSb, with the V-VIII-XV systems, e.g. VFeSb, for which larger degeneracy is obtained for the former due to convergence of the Γ- and L-pockets, despite the lower covalency. A similar plot colored with the maximum calculated power factor is found in Figure S7 in the SI.

Not only the thermoelectric power factor is strongly dependent on the covalent interactions in HHs. Also the response properties, here exemplified by the Born effective charge (BEC), which measures the charge contributing to the polarization during atomic displacement,[39] depend on the covalency. When strong covalent interactions are present between polyanion and cation, anomalously large negative BECs are found on the late transition metal (Figure 3b). Approaching the Zintl picture generally leads to smaller BECs. Opposite to the effect on thermoelectric transport, the BEC is here increased with both increasing XY and YZ covalency. This is an interesting example of the effect of covalency on BECs, which, unlike static charge measures, often increase in magnitude with increasing covalency, especially when the covalency arise from mixing with formally empty d-states. This results in linear relationships between the BEC of the Y atom and the bond order of the XY interaction as well as with the population of the $t_{2g}$ states on the X cation (Figure S8 in the SI). The role of mixing with formally empty d-states is similar to the famous case of perovskite $BaTiO_3$ in the vicinity of a ferroelectric transition.[40] Interestingly, VFeSb, with one of the largest BECs for the Y-site, has previously been identified as a potential piezoelectric material with a similar piezoelectric response to $PbTiO_3$, which forms part of state-of-the-art piezoelectric materials.[41]

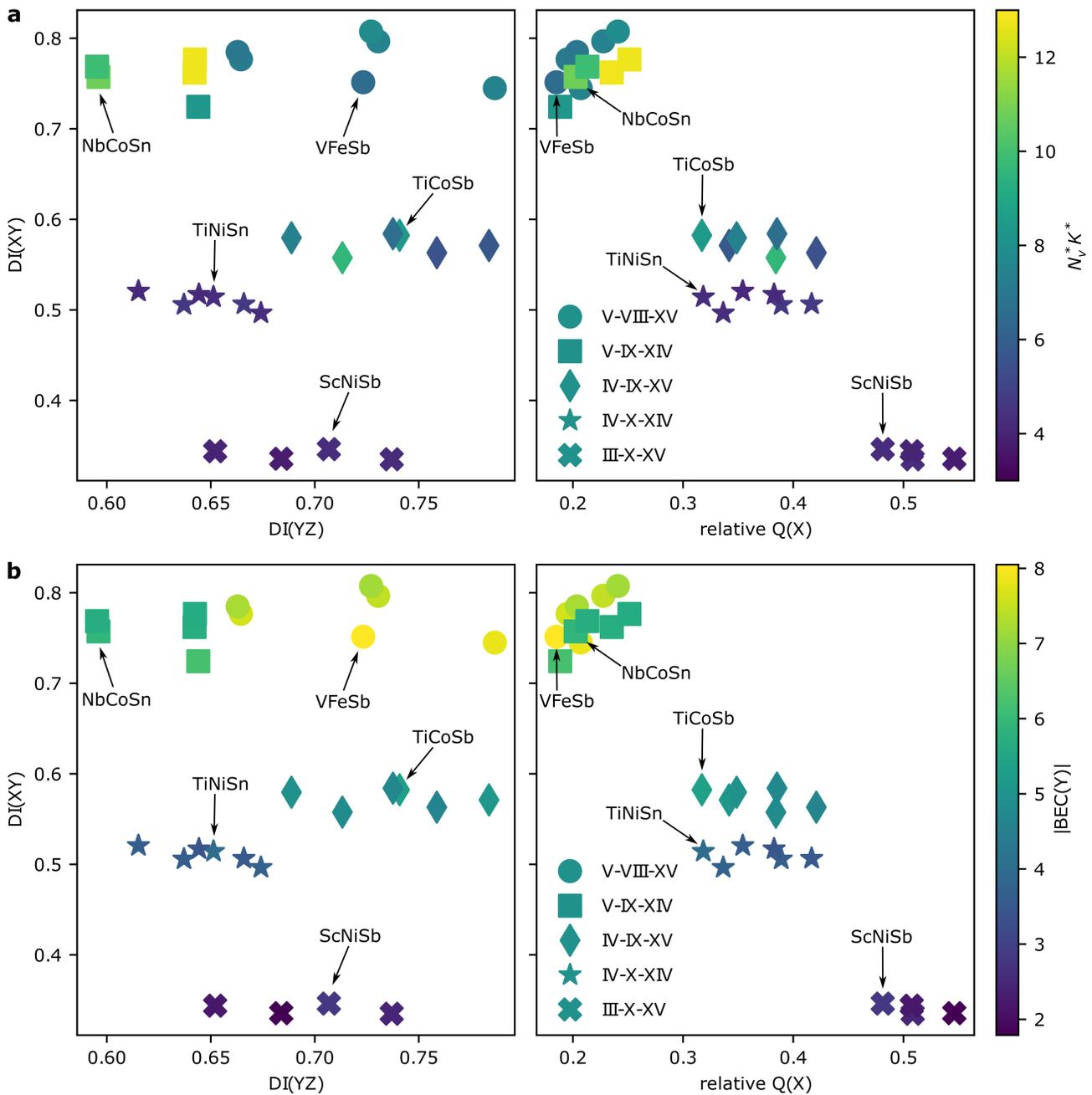

**Figure 3.** Bonding and properties in Half Heuslers. Transition metal containing HHs are marked according to the chemical bonding descriptors as in Figure 2, and colored with their calculated properties. **a** The HHs are colored with their Fermi surface complexity factor evaluated at their maximum power factor for p-type conduction at 700 K, measuring the effective degeneracy times the anisotropy factor for the band structure. This determines the band structure contribution to the thermoelectric power factor. **b** The HHs are colored with the Born effective charge (BEC) of the Y-atom, i.e. the late transition metal residing on the site, which is part of both zinc blende networks in the structure. Numerical values are found in Table S1 and S2 in the SI.

**2.3 Defect chemistry and the Zintl concept in Half-Heuslers**

Optimizing thermoelectric performance relies heavily on tailoring the carrier concentration, making defect chemistry highly important for thermoelectric materials. This includes both the intrinsic defect chemistry and effect of extrinsic dopants.[42] Regarding HHs, the intrinsic defect chemistry of XNiSn systems is very interesting, since they have been observed to form large off-stoichiometries with 2-6 % Ni on the interstitial tetrahedral site (often termed the Full-Heusler site).[26, 27] This has been explained from a subdivision of oxidation states within the Zintl concept, suggesting that the interstitial Ni is in oxidation state zero resulting in a $d^{10}$ configuration and the formation of a stable neutral defect.[13] The defect chemistry of interstitial Ni in XNiSn has been the subject of several investigations, where DFT with GGA(+U) functionals suggests a neutral charge state, and hybrid functionals suggest a defect in the +1 charge state as the most common one,[43, 44] while also clustering of defects has been shown to lead to stabilization.[45]

Resorting to real space chemical bonding analysis, an interesting perspective on the defect chemistry arises. In Figure 4, we compare the deformation densities obtained in pristine TiNiSn and pristine $TiNi_2Sn$ in the Full-Heusler structure with that obtained around an interstitial Ni in the TiNiSn structure. As evident from Figure 4a-c, there is large difference in the deformation electron density around Ni in the zinc blende framework, and the interstitial Ni. The framework Ni shows significant charge accumulation into the Ni-Ti bond, similar to the case of TiCoSb in Figure 1b, whereas the interstitial Ni only shows comparably weak charge redistribution. This is highly similar to the deformation density in $TiNi_2Sn$ shown in Figure 4d. This is also evident from topological analysis of the electron density, since the interstitial Ni has an integrated charge of -0.48e compared to -1.07e for the framework Ni and -0.64e in $TiNi_2Sn$. The electron density at the bond critical points (BCP) are reduced from 0.369 and 0.319 eÅ$^{-3}$ in the Half-Heusler framework to 0.306 and 0.263 eÅ$^{-3}$ around the interstitial, and 0.307 and 0.262 eÅ$^{-3}$ for the Full-Heusler for the Ni-Sn and Ni-Ti bonds, respectively.

Thus, the topological bonding descriptors further corroborate that interstitial Ni has significantly weaker interaction with the framework than the interactions within the framework structure, and interstitial Ni approaches the bonding situation found in the Full-Heusler. Thus, the large stability of interstitial Ni cannot be inferred from the bonding and properties of the framework Ni as suggested based on Zintl chemistry,[13] but rather from the fact that the chemical bonding environment is similar to another stable phase ($TiNi_2Sn$). This also fits well with the observation of clustering of Ni interstitials, forming Full-Heusler like inclusions in the HH matrix.[45]

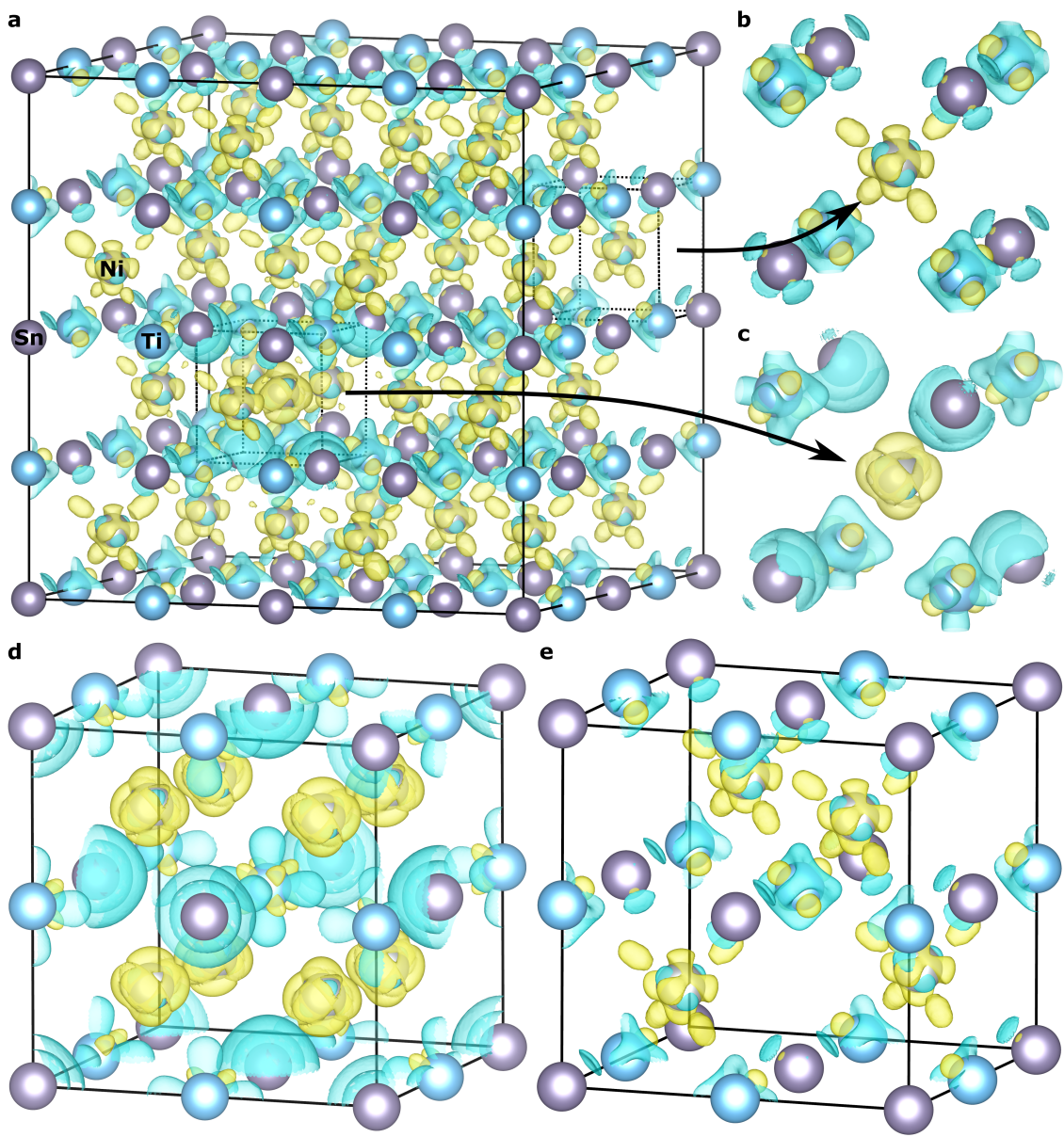

**Figure 4.** Deformation density around interstitial Ni in TiNiSn. Deformation densities of **a** a 2x2x2 supercell of the conventional F-centered unit cell of TiNiSn with one Ni interstitial, with **b** a zoom on a Ni belonging to the HH framework far from the defect, and **c** a zoom on the interstitial Ni. **d** and **e** shows the deformation densities in full-Heusler TiNi$_2$Sn and HH TiNiSn, respectively. Positive isosurfaces are shown in yellow and negative isosurface in blue. The isosurfaces are drawn at 0.01 e au$^{-3}$.

## 4. Conclusions

In summary, we have shown that the chemical bonding in thermoelectric HH semiconductors cannot be inferred from the Zintl concept, and that this has great influence on their transport and response properties. HHs consisting purely of main group elements are reasonably described with Zintl chemistry, allowing for a significant bond polarity within the extended Zintl anion. For those containing transition metals, Zintl chemistry is approached for the largest electronegativity differences between transition metals, but in all cases significant covalent bonding is observed between formal cation and anion. This has significant impact on their band structure and thermoelectric properties, where increased covalency between formal cation and anion, and therefore larger deviations from Zintl chemistry, stabilizes the highest occupied states at the Γ-point relative to the L- and W-points, leading to increased band degeneracy and improved p-type thermoelectric properties. Similarly, also the defect chemistry is more complex than what can be inferred from Zintl chemistry, particularly in the case of XNiSn, where the stability of interstitial Ni is related to the similar bonding network in stable Full-Heusler structures, rather than to the properties inferred from the framework Ni atoms. In broader terms, the detailed analysis of chemical bonding and its relation to physical properties paves the way for better understanding of complex materials, including thermoelectrics, for which current design strategies and interpretations often are based on too simple chemical concepts.

## 5. Methods

### 5.1 Theoretical calculations

Density functional theory calculation were performed on 27 transition metal containing Half-Heuslers and three main group HHs (LiSiAl, LiAsZn, LiAsMg). The calculations were performed using the plane wave pseudopotential method as implemented in QuantumEspresso (QE).[46, 47] Projector augmented wave (PAW) pseudopotentials from the PSL-library were used for all atoms.[48] Calculations were performed within the generalized gradient approximation (GGA) with the Perdew-Burke-Erzerhof (PBE) exchange-correlation functional.[49] An energy cut-off of 100 Ry and a 10x10x10 Γ-centered k-point grid in reciprocal space were used for the calculations, and the lattice parameters were optimized.

Born effective charges were calculated using density functional perturbation theory as implemented in the Phonon module of QE.

Analysis of chemical bonding through the resulting electron densities within the QTAIM[29] framework was performed with CRITIC2.[50] The reconstructed PAW density was used to determine the atomic basins and for determination of electron density topology, and the valence electron density was used as the integrand in the atomic basins. Calculations of delocalization indices[30, 31] (DI) using maximally localized Wannier functions[51] were also performed in CRITIC2 using the output wave function from QE.[52] Due to the large required computational resources for the DI calculations, only 6x6x6 k-point grids were used for these, and as they require the use of norm-conserving pseudopotentials, the database of optimized norm-conserving Vanderbilt pseudopotentials were used.[53] Thermoelectric transport properties were calculated with BoltzTraP interfaced with QE,[54] and Fermi surface complexity factor was calculated using the methods described by Gibbs *et al.*[38]

Supercell calculations were performed on a 2x2x2 supercell of the conventional unit cell of TiNiSn (96 atoms in the pristine system) with one additional Ni added at the interstitial site. Atomic positions were relaxed, while lattice parameters were kept fixed.

**Acknowledgements**

This work was supported by the Villum Foundation. Affiliation with the Aarhus University Center for Integrated Materials Research (iMAT) is gratefully acknowledged. The theoretical calculation were performed at the Centre for Scientific Computing, Aarhus.

# Supporting information for

# Chemical bonding origin of the thermoelectric power factor in Half-Heusler semiconductors

Kasper Tolborg[a] and Bo B. Iversen[a]*

a) Center for Materials Crystallography, Department of Chemistry and iNANO, Aarhus University, Langelandsgade 140, 8000 Aarhus C, Denmark
Corresponding author: bo@chem.au.dk

Table S1. Chemical bonding descriptors for Half-Heuslers. The unit cell parameters, integrated QTAIM charges, Q, delocalization indices, DI, between nearest neighbours of each pair, density at the bond critical point, $\rho_{BCP}$, for the found BCPs, and d-orbital populations, d, for the $t_{2g}$ and $e_g$-orbitals for the X-atom are given in the table.

| Formula | Unit cell [a.u.] | Q(X) | Q(Y) | Q(Z) | DI(XY) | DI(YZ) | DI(XZ) | $\rho_{BCP}$(XY) [eÅ$^{-3}$] | $\rho_{BCP}$(YZ) [eÅ$^{-3}$] | $d_{t2g}$(X) | $d_{eg}$(X) |
|---|---|---|---|---|---|---|---|---|---|---|---|
| HfCoSb | 11.4430 | 1.54 | -1.01 | -0.53 | 0.558 | 0.713 | 0.254 | 0.372 | 0.367 | 1.65 | 0.86 |
| HfNiSn | 11.5373 | 1.56 | -1.19 | -0.36 | 0.506 | 0.637 | 0.249 | 0.337 | 0.337 | 1.46 | 0.94 |
| HfPdSn | 12.0000 | 1.67 | -1.51 | -0.15 | 0.507 | 0.666 | 0.215 | 0.342 | 0.336 | 1.57 | 0.93 |
| HfRhSb | 11.8888 | 1.68 | -1.41 | -0.27 | 0.563 | 0.759 | 0.212 | 0.384 | 0.376 | 1.76 | 0.83 |
| NbCoSn | 11.2770 | 1.01 | -0.94 | -0.08 | 0.757 | 0.596 | 0.300 | 0.416 | 0.363 | 2.36 | 1.20 |
| NbFeSb | 11.2706 | 0.98 | -0.65 | -0.33 | 0.777 | 0.665 | 0.318 | 0.435 | 0.387 | 2.57 | 1.09 |
| NbRhSn | 11.7136 | 1.18 | -1.32 | 0.15 | 0.762 | 0.642 | 0.249 | 0.432 | 0.372 | 2.50 | 1.19 |
| NbRuSb | 11.7034 | 1.14 | -1.05 | -0.09 | 0.797 | 0.731 | 0.261 | 0.455 | 0.405 | 2.71 | 1.05 |
| ScNiSb | 11.5516 | 1.44 | -0.82 | -0.63 | 0.346 | 0.707 | 0.193 | 0.259 | 0.353 | 1.28 | 0.76 |
| ScPdSb | 12.0500 | 1.53 | -1.12 | -0.41 | 0.335 | 0.737 | 0.168 | 0.257 | 0.351 | 1.00 | 0.54 |
| TaCoSn | 11.2606 | 1.07 | -1.01 | -0.05 | 0.769 | 0.595 | 0.306 | 0.438 | 0.364 | 2.19 | 1.17 |
| TaFeSb | 11.2549 | 1.02 | -0.73 | -0.29 | 0.785 | 0.663 | 0.326 | 0.458 | 0.388 | 2.41 | 1.08 |
| TaRhSn | 11.6996 | 1.26 | -1.41 | 0.16 | 0.777 | 0.642 | 0.255 | 0.454 | 0.373 | 2.34 | 1.16 |
| TaRuSb | 11.6910 | 1.20 | -1.14 | -0.07 | 0.807 | 0.727 | 0.268 | 0.478 | 0.404 | 2.55 | 1.04 |
| TiCoSb | 11.1108 | 1.27 | -0.89 | -0.38 | 0.582 | 0.741 | 0.238 | 0.355 | 0.411 | 1.43 | 0.54 |
| TiNiSn | 11.2232 | 1.27 | -1.07 | -0.20 | 0.514 | 0.651 | 0.237 | 0.319 | 0.369 | 1.57 | 0.89 |
| TiPdSn | 11.7452 | 1.35 | -1.37 | 0.03 | 0.497 | 0.674 | 0.205 | 0.314 | 0.363 | 1.30 | 0.68 |
| TiRhSb | 11.6135 | 1.37 | -1.26 | -0.10 | 0.571 | 0.784 | 0.200 | 0.357 | 0.416 | 1.49 | 0.52 |
| VCoSn | 10.9851 | 0.94 | -0.95 | 0.01 | 0.724 | 0.644 | 0.265 | 0.382 | 0.397 | 2.41 | 1.11 |
| VFeSb | 10.9648 | 0.93 | -0.67 | -0.26 | 0.751 | 0.723 | 0.279 | 0.401 | 0.431 | 2.62 | 0.96 |
| VRuSb | 11.4284 | 1.04 | -1.05 | 0.02 | 0.745 | 0.787 | 0.226 | 0.410 | 0.449 | 2.73 | 0.92 |
| YNiSb | 12.0040 | 1.52 | -0.76 | -0.77 | 0.344 | 0.652 | 0.210 | 0.252 | 0.306 | 1.41 | 0.92 |
| YPdSb | 12.4811 | 1.64 | -1.05 | -0.59 | 0.336 | 0.684 | 0.181 | 0.254 | 0.304 | 1.06 | 0.64 |
| ZrCoSb | 11.5208 | 1.39 | -0.87 | -0.52 | 0.580 | 0.689 | 0.263 | 0.360 | 0.358 | 1.77 | 0.86 |
| ZrNiSn | 11.5892 | 1.42 | -1.06 | -0.35 | 0.521 | 0.615 | 0.259 | 0.329 | 0.332 | 1.58 | 0.97 |
| ZrPdSn | 12.0748 | 1.53 | -1.38 | -0.16 | 0.517 | 0.644 | 0.224 | 0.332 | 0.328 | 1.67 | 0.96 |
| ZrRhSb | 11.9562 | 1.54 | -1.27 | -0.28 | 0.584 | 0.738 | 0.220 | 0.375 | 0.369 | 1.88 | 0.84 |
| LiSiAl | 11.2146 | 0.84 | -2.42 | 1.58 | 0.069 | 0.488 | 0.008 | 0.110 | 0.321 | - | - |
| LiAsMg | 11.7116 | 0.85 | -2.37 | 1.52 | 0.076 | 0.223 | 0.003 | 0.100 | 0.187 | - | - |
| LiAsZn | 11.2587 | 0.85 | -1.25 | 0.40 | 0.060 | 0.617 | 0.009 | 0.110 | 0.345 | - | - |

**Table S2.** Response properties and thermoelectric properties for Half-Heuslers. Born effective charges, BEC, the energy difference between the highest occupied level at the Γ- and L-points, the maximum p-type thermoelectric power factor divided by the relaxation time, $(S^2\sigma/\tau)_{max}$, the conductivity, $m_c^*$, and Seebeck effective mass, $m_S^*$, and the Fermi surface complexity factor, $N_V^*K^*$, are given in the table. The effective masses and Fermi surface complexity are evaluated for the maximum power factor. Thermoelectric properties are evaluated at 700 K.

| Formula | BEC(X) | BEC(Y) | BEC(Z) | $\Delta E_{\Gamma L}$ [eV] | $(S^2\sigma/\tau)_{max}$ [$10^{14}$ μW cm$^{-1}$ K$^{-2}$ s$^{-1}$] | $m_c^*$ [$m_e$] | $m_S^*$ [$m_e$] | $N_V^*K^*$ |
|---|---|---|---|---|---|---|---|---|
| HfCoSb | 2.91 | -4.82 | 1.91 | -0.02 | 173.63 | 1.55 | 6.99 | 9.60 |
| HfNiSn | 2.72 | -3.64 | 0.91 | 0.55 | 63.45 | 0.84 | 2.34 | 4.62 |
| HfPdSn | 2.77 | -3.52 | 0.75 | 0.76 | 58.69 | 0.70 | 1.96 | 4.67 |
| HfRhSb | 3.08 | -4.64 | 1.55 | 0.49 | 67.87 | 0.98 | 3.06 | 5.51 |
| NbCoSn | 4.25 | -5.88 | 1.63 | -0.34 | 186.19 | 1.38 | 6.72 | 10.78 |
| NbFeSb | 5.05 | -7.62 | 2.56 | -0.89 | 117.81 | 1.19 | 4.42 | 7.16 |
| NbRhSn | 4.33 | -5.71 | 1.36 | -0.06 | 188.21 | 0.94 | 5.16 | 12.80 |
| NbRuSb | 5.16 | -7.47 | 2.31 | -0.49 | 102.22 | 0.83 | 3.15 | 7.41 |
| ScNiSb | 2.10 | -2.77 | 0.67 | 0.84 | 45.64 | 0.51 | 1.33 | 4.26 |
| ScPdSb | 2.22 | -2.43 | 0.20 | 1.12 | 41.31 | 0.42 | 1.09 | 4.24 |
| TaCoSn | 4.00 | -5.68 | 1.69 | -0.24 | 181.82 | 1.55 | 7.10 | 9.84 |
| TaFeSb | 4.60 | -7.22 | 2.62 | -0.79 | 114.47 | 1.19 | 4.34 | 6.95 |
| TaRhSn | 4.12 | -5.63 | 1.51 | 0.05 | 216.89 | 1.35 | 7.36 | 12.74 |
| TaRuSb | 4.76 | -7.18 | 2.42 | -0.35 | 108.54 | 1.03 | 4.06 | 7.84 |
| TiCoSb | 3.21 | -5.35 | 2.14 | 0.05 | 167.67 | 1.87 | 7.76 | 8.42 |
| TiNiSn | 2.83 | -3.97 | 1.14 | 0.41 | 79.24 | 1.49 | 3.95 | 4.31 |
| TiPdSn | 2.78 | -3.73 | 0.97 | 0.49 | 76.61 | 1.39 | 3.67 | 4.31 |
| TiRhSb | 3.20 | -5.11 | 1.91 | 0.33 | 96.04 | 1.39 | 4.45 | 5.71 |
| VCoSn | 4.54 | -6.20 | 1.66 | -0.15 | 178.36 | 2.19 | 8.97 | 8.30 |
| VFeSb | 5.38 | -8.05 | 2.67 | -0.52 | 121.52 | 1.53 | 5.34 | 6.50 |
| VRuSb | 5.43 | -7.83 | 2.42 | -0.27 | 130.25 | 1.37 | 5.23 | 7.46 |
| YNiSb | 2.02 | -2.20 | 0.19 | 0.71 | 39.45 | 0.43 | 1.08 | 4.01 |
| YPdSb | 2.32 | -1.79 | -0.56 | 1.19 | 31.22 | 0.29 | 0.71 | 3.82 |
| ZrCoSb | 3.01 | -4.94 | 1.94 | -0.17 | 146.73 | 1.76 | 6.65 | 7.34 |
| ZrNiSn | 2.65 | -3.64 | 0.98 | 0.41 | 67.12 | 1.15 | 2.97 | 4.15 |
| ZrPdSn | 2.65 | -3.45 | 0.79 | 0.63 | 59.65 | 0.90 | 2.33 | 4.19 |
| ZrRhSb | 3.10 | -4.72 | 1.61 | 0.31 | 97.81 | 1.08 | 3.79 | 6.62 |
| LiSiAl | 1.00 | -2.73 | 1.67 | 1.09 | 54.62 | 0.26 | 0.96 | 7.06 |
| LiAsMg | 0.93 | -2.74 | 1.81 | 0.61 | 71.40 | 0.67 | 2.16 | 5.80 |
| LiAsZn | 1.04 | -3.14 | 2.05 | 1.04 | 52.69 | 0.37 | 1.19 | 5.74 |

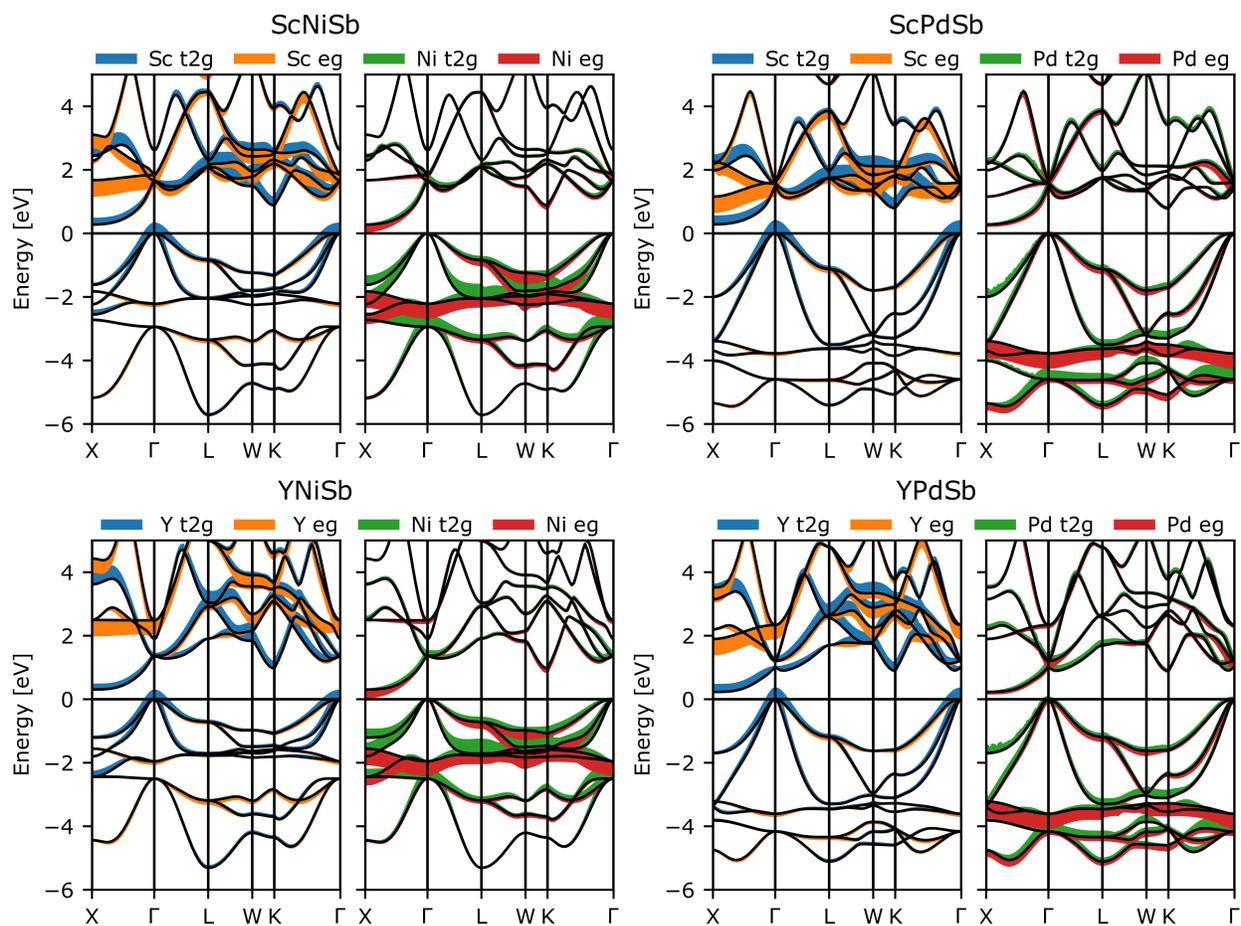

**Figure S1.** Fatband plots for group III-X-XV Half-Heuslers. Band structures with the states projected onto d-orbitals of the transition metals. The thickness of the colored band indices the weight of the orbital for the given band at the given k-point.

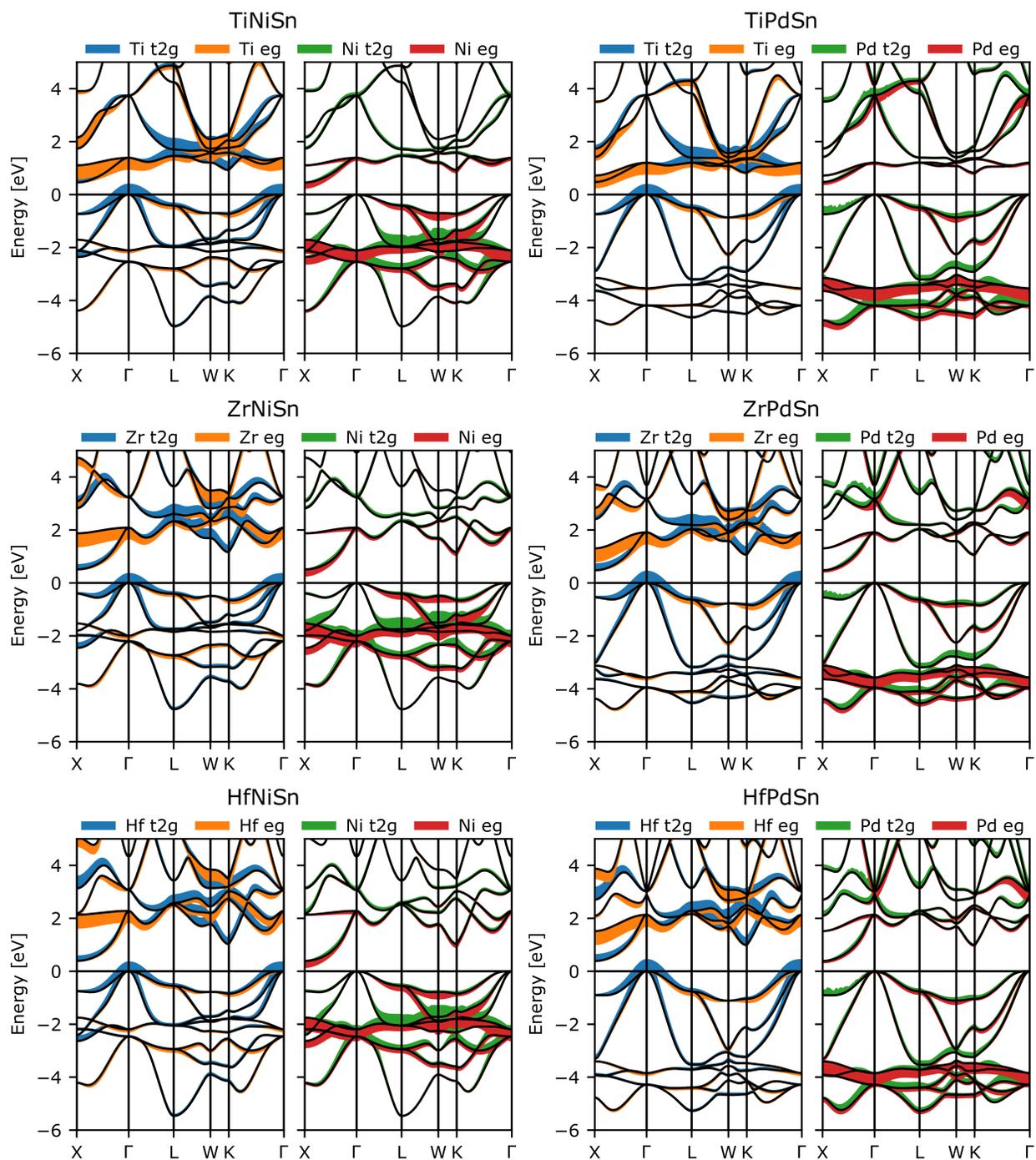

**Figure S2.** Fatband plots for group IV-X-XIV Half-Heuslers. Band structures with the states projected onto d-orbitals of the transition metals. The thickness of the colored band indices the weight of the orbital for the given band at the given k-point.

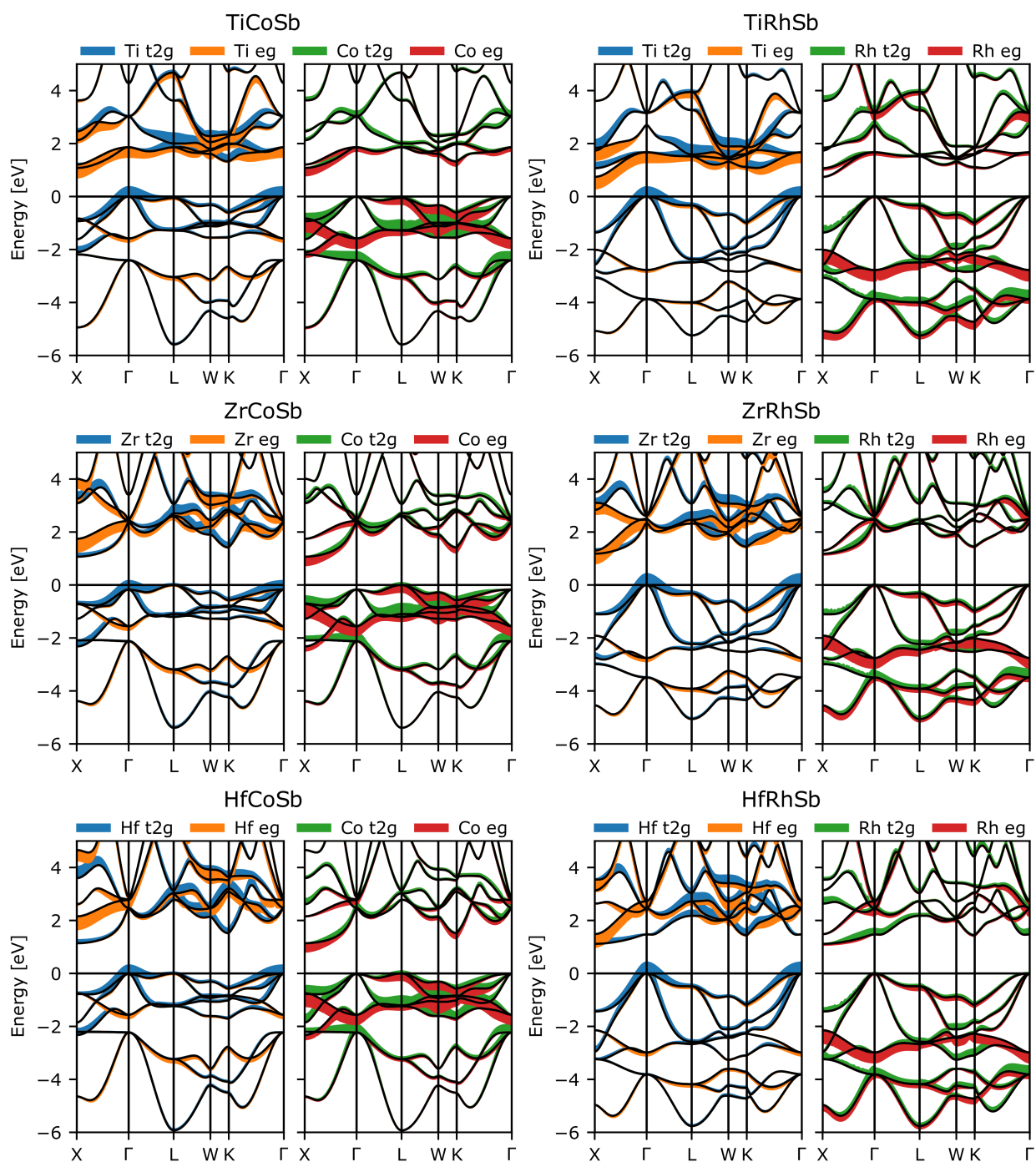

**Figure S3.** Fatband plots for group IV-IX-XV Half-Heuslers. Band structures with the states projected onto d-orbitals of the transition metals. The thickness of the colored band indices the weight of the orbital for the given band at the given k-point.

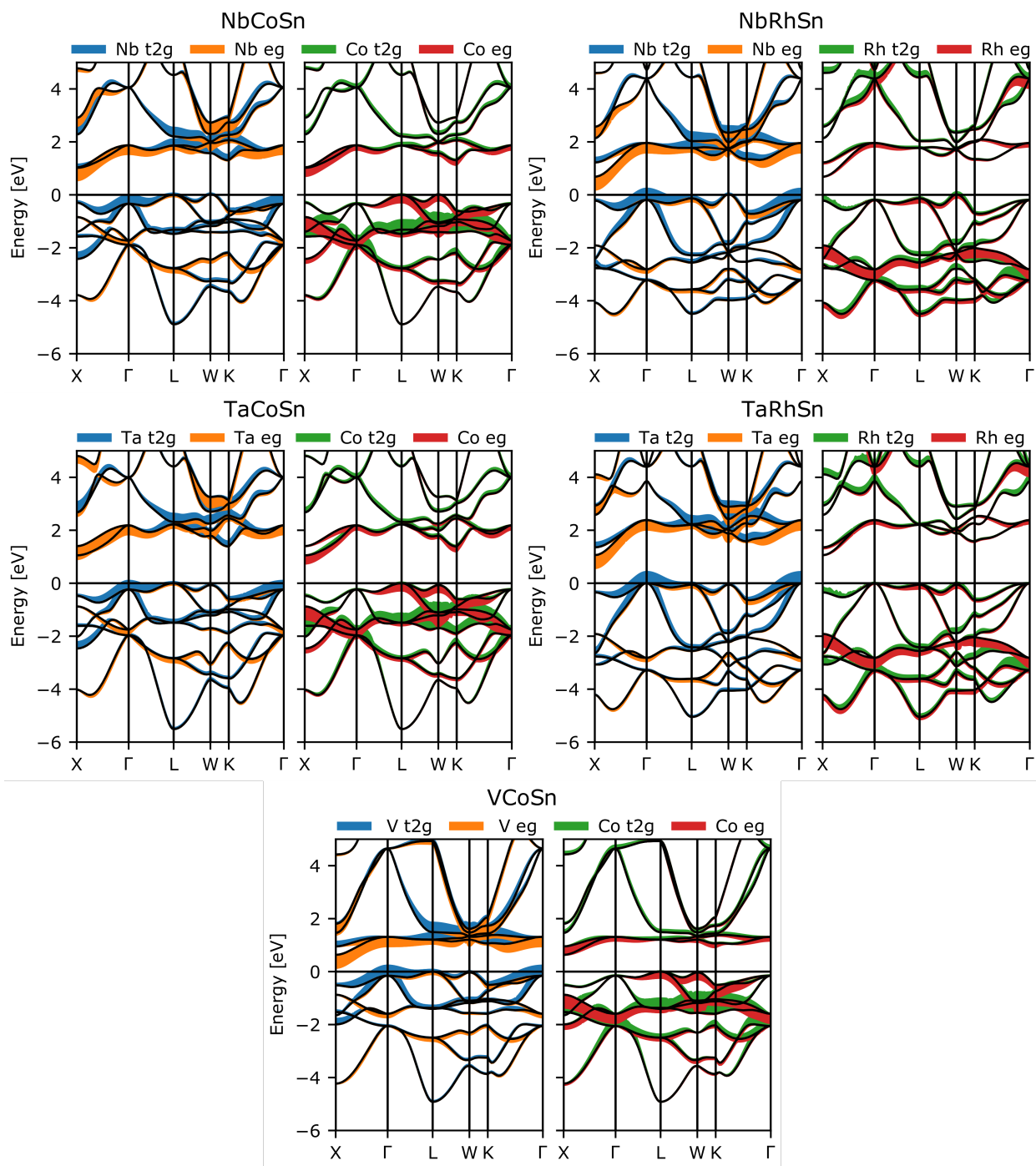

**Figure S4.** Fatband plots for group V-IX-XIV Half-Heuslers. Band structures with the states projected onto d-orbitals of the transition metals. The thickness of the colored band indices the weight of the orbital for the given band at the given k-point.

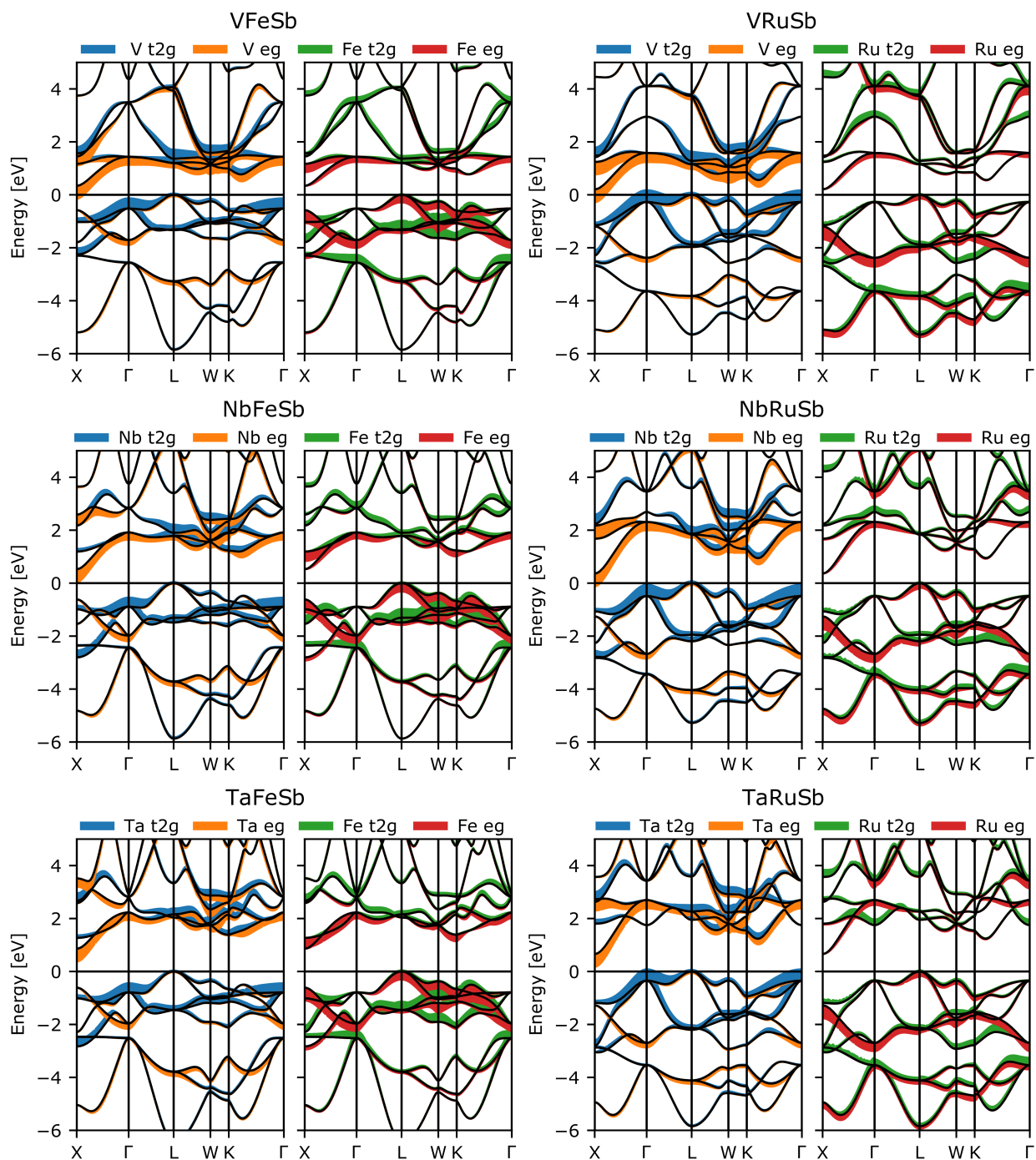

**Figure S5.** Fatband plots for group V-VIII-XV Half-Heuslers. Band structures with the states projected onto d-orbitals of the transition metals. The thickness of the colored band indices the weight of the orbital for the given band at the given k-point.

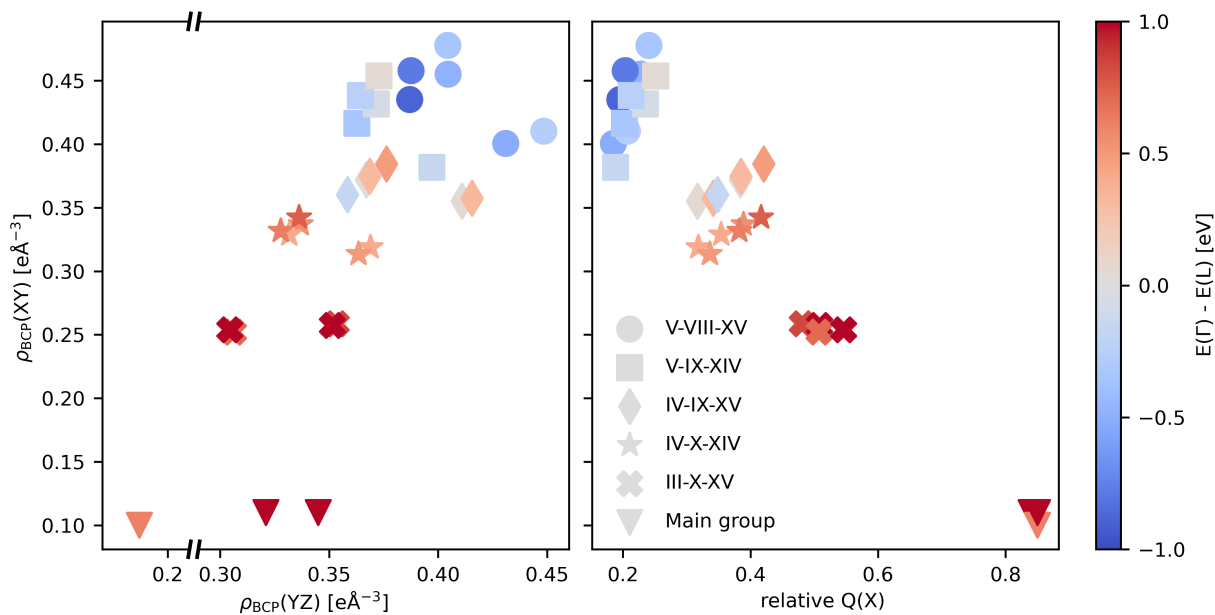

**Figure S6.** Maps of chemical bonding in Half Heuslers. Main group and transition metal containing HHs are marked according to their chemical bonding descriptors. On the ordinate the density at the bond critical point (BCP) between X and Y atoms is shown, and on the abscissa the density at the BCP between Y and Z is shown to left, and the charge transfer from X to YZ divided by the formal oxidation state is shown to the right. The materials containing elements from the same groups in the periodic table are marked with the same symbols for the transition metal containing ones, and with one symbol for all main group HHs. The materials are colored according to the energy difference between the highest occupied states at the Γ-point and L-point in the first Brillouin zone. Note that the observable electron density shows very similar behavior to the non-observable delocalization indices from Figure 2 in the main article, although the divisions between groups are slightly weaker here.

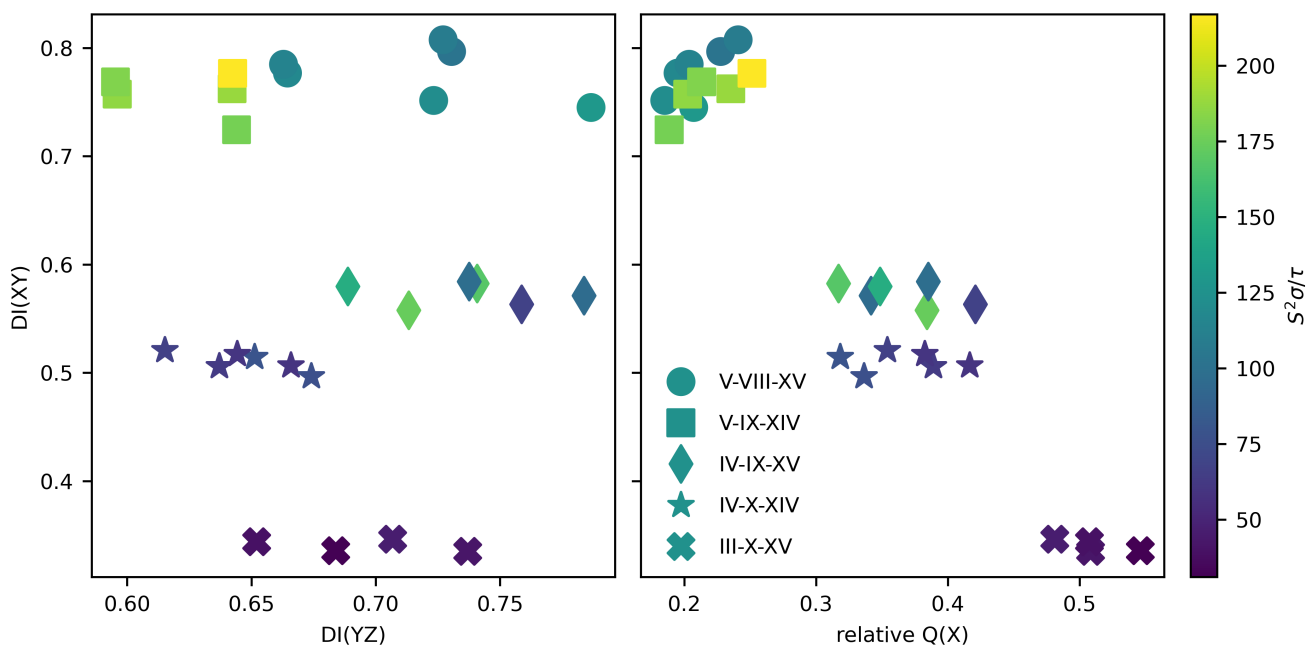

**Figure S7.** Bonding and thermoelectric power factor. Transition metal containing HHs are marked according to the chemical bonding descriptors as in Figure 2 in the main article, and colored with their calculated maximum power factor within the constant relaxation time approximation divided by the relaxation time to make the values independent of relaxation time. $S^2\sigma/\tau$ is given in units of $10^{14}$ µW cm$^{-1}$ K$^{-2}$ s$^{-1}$.

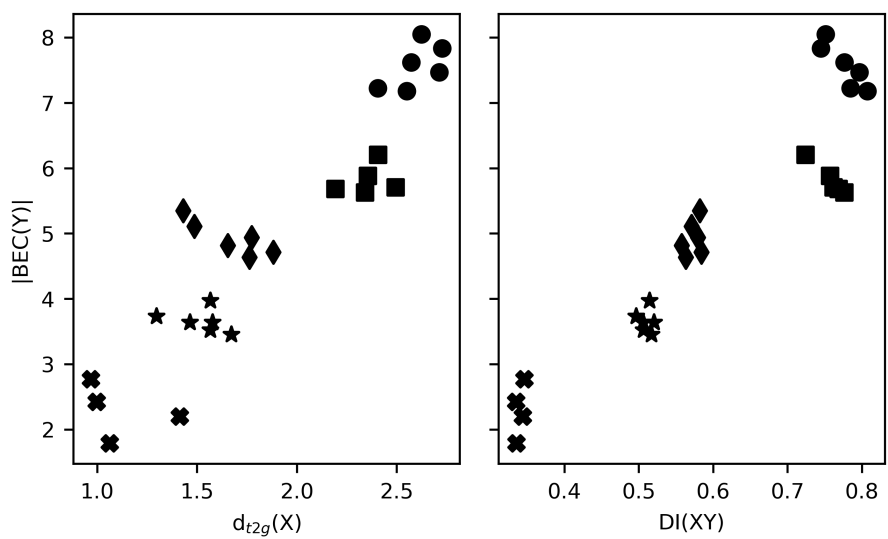

**Figure S8.** Relation between bonding and response. Absolute value of the Born effective charge (BEC) of the Y atom as a function of the population of the $t_{2g}$-orbitals on X and the delocalization index between X and Y. The increased population of the $t_{2g}$-orbitals on X is a result of the overlap between d-orbitals on X and Y, and therefore an orbital based measure of the orbital mixing or covalency of the XY interaction. The markers symbolize materials from different groups, and their interpretation is given in Figure 2 in the main article.